\documentstyle[12pt]{article}
\begin{document}
\pagestyle{empty}

\begin{titlepage}
\rightline{IC/98/40}
\vspace{2.0 truecm}
\begin{center}
\begin{Large}
{\bf Model-independent limits on four-fermion contact interactions at LC with 
polarization}

\end{Large}
\end{center}
\vspace{0.5cm}
\begin{center}
{\large  A. A. Pankov\hskip 2pt\footnote{Permanent address: 
Department of Physics, Polytechnical Institute, Gomel, 
246746 Belarus. E-mail: PANKOV@GPI.GOMEL.BY}
}\\[0.3cm]
The Abdus Salam International Centre for Theoretical Physics, Trieste, Italy, 
\\
Istituto Nazionale di Fisica Nucleare, Sezione di Trieste, Trieste, Italy

\vspace{5mm}

{\large  N. Paver\hskip 2pt\footnote{Also supported by the Italian
Ministry of University, Scientific Research and Technology (MURST).}
}\\[0.3cm]
Dipartimento di Fisica Teorica, Universit\`{a} di Trieste,
Trieste, Italy\\
Istituto Nazionale di Fisica Nucleare, Sezione di Trieste, Trieste,
Italy
\end{center}
\vspace{2.0cm}

\begin{abstract}
\noindent
Fermion compositeness, and other types of new physics that can be described 
by the exchange of very massive particles, can manifest themselves as the 
result of an effective four-fermion contact interaction.
In the case of the processes 
$e^+e^-\to \mu^+\mu^-,\tau^+\tau^-,\bar{b}b$ and $\bar{c}c$ at future $e^+e^-$ 
colliders with $\sqrt{s}=0.5-1$ TeV, we examine the sensitivity to four-fermion 
contact interactions of two new integrated observables, $\sigma_+$ and 
$\sigma_-$, conveniently defined for such kind of analysis. We find that, if 
longitudinal polarization of the electron beam were available, these 
observables would offer the opportunity to separate the 
helicity cross sections and, in this way, to derive model-independent 
bounds on the relevant parameters. 
 
\vspace*{3.0mm}

\noindent
\end{abstract}
\end{titlepage}

\pagestyle{plain}
\setlength{\baselineskip}{1.3\baselineskip}

\section{Introduction}

Deviations from the Standard Model (SM) caused by new physics characterized 
by very high mass scales $\Lambda$ can systematically be studied at lower
energies by using the effective Lagrangian approach. In this framework, 
by integration of the heavy degrees of freedom of the new  
theory, an effective Lagrangian which obeys the low energy SM symmetries is 
constructed in terms of the SM fields. The resulting interaction consists of 
the SM itself as the leading term, plus a series of higher order terms 
represented by higher-dimensional local operators that are suppressed by 
powers of the scale $\Lambda$. Consequently, the effects 
of the new physics can be observed at energies well-below $\Lambda$ as a 
deviations from the SM predictions, and can be related to some effective 
contact interaction. Here, we study the manifestations of such four-fermion 
contact interactions \cite{Eichten}-\cite{Ruckl} in high-energy $e^+e^-$ 
collisions.
\par
In the context of composite models of leptons and quarks, the contact 
interaction is regarded as a remnant of the binding force between the 
fermion substructure constituents. Furthermore, in $e^+e^-$ collisions, 
many types of new physics, for which the exchanged particles in the $s$, $t$, 
or $u$ channels have mass-squared much larger than the corresponding 
Mandelstam invariant variables, can be described by an effective $eeff$ 
contact term in the interaction Lagrangian \cite{Schrempp}-\cite{Zerwas1}.
For example, effects of a $Z^\prime$ boson of a few TeV mass scale would be 
well-represented by a four-fermion contact interaction. The exchange of a 
leptoquark of a similar mass scale could be described by an effective 
$eeqq$ contact term in the relevant interaction. At energies much lower than 
the sparticle masses, $R$-parity breaking interactions introduce effective 
$eell$ and $eeqq$ interactions. The concept of contact interactions with a 
universal energy scale $\Lambda$ is also used in other processes, such as 
$ep$ and $p\bar{p}$ collisions, to search for substructure of quarks or new 
heavy particles coupling to quarks and gluons. Thus, quite generally, the 
contact interaction is considered as a convenient parameterisation of 
deviations from the SM that may be caused by some new physics at the large
scale~$\Lambda$. 
\par
Fermion-pair production in $e^+e^-$ collisions 
\begin{equation}
e^++e^-\to \bar{f}+f \label{proc}
\end{equation}
($f=l$ or $q$) is one of the basic processes of the SM, and deviations of the 
measured observables from the predicted values would be a first indication of 
new physics beyond the SM.
\par
The lowest order four-fermion contact terms have dimension 6, which implies 
that they are suppressed by $g^2_{eff}/\Lambda^2$, with $g_{eff}$ an effective 
coupling constant. The fermion currents are restricted to be helicity 
conserving and flavor diagonal. The general, $SU(3)\times SU(2)\times U(1)$ 
invariant, contact four-fermion $eeff$ interaction Lagrangian with dimension 6 
can be written as \cite{Eichten}-\cite{Ruckl}, \cite{Godfrey}-\cite{Barger2}:
\begin{eqnarray}
{{\cal L}}=\frac{g^2_{eff}}{2\Lambda^2}\left[
\eta_{LL}\left(\bar e_L\gamma_\mu e_L\right)\left(\bar f_L\gamma^\mu f_L\right) 
+\eta_{LR}\left(\bar{e}_L\gamma_\mu e_L\right)\left(\bar f_R\gamma^\mu f_R
\right)\right. \nonumber \\
+\left.\eta_{RL}\left(\bar{e}_R\gamma_\mu e_R\right)\left(\bar f_L
\gamma^\mu f_L\right) +
\eta_{RR}\left(\bar e_R\gamma_\mu e_R\right)\left(\bar f_R\gamma^\mu f_R\right) 
\right],
\label{lagra}
\end{eqnarray}
where generation and color indices have been suppressed. The subscripts 
$L,R$ indicate that the current in each parenthesis can be either left- or 
right-handed, and the parameters $\eta_{\alpha\beta}$ ($\alpha,\beta=R,L$) 
determine the chiral structure of the interaction. They are free parameters 
in these models, but typical values are between $-1$ and $+1$, depending 
on the type of the assumed theory \cite{Schrempp}. 
It is conventional to define $g^2_{eff}=4\pi$, and the interaction is 
defined to be strong when $\sqrt s$ approaches $\Lambda$. 
\par 
Constraints on the parameters characterizing contact interactions can be 
derived phenomenologically, by comparing the SM prediction for the 
observables that can involve such interactions with the relevant experimental
data. In principle, cross sections simultaneously depend on all four-fermion 
effective coupling constants in Eq.~(\ref{lagra}), which therefore cannot be 
easily disentangled 
{\it a priori}. Therefore, such analyses are usually performed by assuming  
a non-zero value for only one parameter at a time, and all the remaining ones 
to vanish. By this procedure, limits on $eeqq$ contact interaction parameters 
have recently been derived from a global analysis of the data \cite{Barger2}, 
including deep inelastic scattering from ZEUS and H1, atomic physics parity 
violation in Cesium from JILA, scattering of polarized $e^-$ on nuclei 
at SLAC, Drell-Yan production at the Tevatron, and the $e^+e^-$ total cross 
section into hadrons at LEP. The obtained constraints exclude, at the 
$95\%$ C.L., contact interactions among leptons and {\it up} or  
{\it down} quarks with scale $\Lambda<7.2-15.4$ TeV. The analogous bounds 
from LEP2 at $\sqrt s=130-183$ GeV on the scale $\Lambda$ of the four-lepton 
contact interactions $eell$ are in the range $5.2-6.5$ TeV, 
depending on the various helicity combinations \cite{aleph}-\cite{delphi}. 
Also, the limits on lepton-heavy quark ($eebb$) contact parameters derived 
from LEP2 are slightly less stringent, and the limits on the corresponding 
mass scale are $1.8-5.9$ TeV \cite{aleph}. Compositeness scales of 
$3.0-6.3$ TeV have been probed at the Tevatron \cite{tevatron}.
\par
In the near perspective, run II of the Tevatron is expected to improve these 
limits to 10 TeV while, in the more distant future, a search limit for 
$\Lambda$ in the range $15-20$ TeV is expected at the LHC \cite{Hinchliffe}.
The next linear $e^+e^-$ collider (LC) with $\sqrt s\geq 500\hskip 2pt GeV$ 
will provide the best opportunities to analyse $eell$, $eebb$ and $eecc$ 
contact interactions with significant accuracy from process (\ref{proc}), due 
to the really high sensitivity of this reaction at such energies, in 
particular if initial beam polarization will be available. A detailed 
analysis of the potential of lepton colliders along these lines was performed 
in \cite{Godfrey} and \cite{Harris}. 
\par
The aim of this paper is to outline an analysis of $eeff$ contact interactions
at the LC with longitudinally polarized beams, on a somewhat different basis
that allows the various effective couplings to be taken into account 
as free parameters simultaneously. To this purpose, we consider two 
particularly helpful integrated observables, $\sigma_+$ and $\sigma_-$, that 
could directly distinguish the relevant helicity cross sections and,
correspondingly, exhibit the dependence on a single 
effective coupling. As the outcome of such a procedure, one should obtain 
disentangled, model-independent, constraints on the $eell$, $eecc$ and 
$eebb$ couplings. In Sec. 2, the separation of the helicity cross sections by
means of the observables $\sigma_+$ and $\sigma_-$ is discussed. In Sec. 3, 
we present the corresponding analysis of the four-fermion couplings, as well 
as the numerical results for the expected bounds at the LC with some conclusive 
remarks.

\section{Separation of the helicity cross sections}

In Born approximation including $\gamma$, $Z$ exchanges and the four-fermion 
contact interaction term (\ref{lagra}), and neglecting $m_f$ with respect 
to the CM energy $\sqrt s$, the differential cross section for the process 
$e^+e^-\to \bar{f} f$ ($f\neq e, t$) with longitudinally polarized 
electron-positron beams can be written as   
\begin{equation}
\frac{d\sigma}{d\cos\theta}
=N_C\,\frac{\pi\alpha_{e.m.}^2}{2s}
\left[(1+\cos^2\theta)\ F_1 +2\cos\theta\ F_2\right], \label{cross} 
\end{equation}
where $\theta$ is the angle between the initial electron and the outgoing
fermion in the CM frame, and $N_C$ is the color factor ($N_C=3$ or $1$ for
final quarks or leptons, respectively). 
The functions $F_{1,2}$ can be expressed in terms of helicity 
amplitudes as
\begin{eqnarray}
F_{1,2}&=&\frac{1}{4}\left[
\left(1+P_e\right)\left(1- P_{\bar{e}}\right)\left(\vert A_{RR}\vert^2
\pm\vert A_{RL}\vert^2\right)\right. \nonumber \\
& &+\left.\left(1-P_e\right)\left(1+P_{\bar{e}}\right)\left(\vert A_{LL}\vert^2
\pm\vert A_{LR}\vert^2\right)\right],
\label{f12} 
\end{eqnarray}
where $P_e$ and ${P_{\bar{e}}}$ are the degrees of longitudinal electron and 
positron polarizations, respectively. The helicity amplitudes 
$A_{\alpha\beta}$ ($\alpha,\beta=L,R$) can be written as
\begin{equation}
A_{\alpha\beta}=(Q_e)_\alpha(Q_f)_\beta+g_\alpha^e\,g_\beta^f\,\chi_Z+
\frac{s\eta_{\alpha\beta}}{2\alpha_{e.m.}\Lambda^2}, 
\label{amplit} 
\end{equation}
where the gauge boson propagator is $\chi_Z=s/(s-M^2_Z+iM_Z\Gamma_Z)$, 
the left- and right-handed fermion couplings are
$g_L^f=(I_{3L}^f-Q_f s_W^2)/s_W c_W$ and $g_R^f=Q_f s_W^2/s_W c_W$ with 
$s_W^2=1-c_W^2\equiv \sin^2\theta_W$, and $Q_f$ are the fermion electric 
charges. 
\par 
Our discussion of the effects of the four-fermion contact interaction in the 
annihilation process (\ref{proc}) will be based on the `new' observables 
$\sigma_{+}$ and $\sigma_-$, defined as the differences of integrated cross 
sections:

\begin{eqnarray}
\label{sigma+}
\sigma_+&\equiv&\left(\int_{-z^*}^1-\int_{-1}^{-z^*}\right)
\frac{d\sigma}{d\cos\theta}\ d\cos\theta, \\
\label{sigma-}
\sigma_-&\equiv&\left(\int_{-1}^{z^*}-\int_{z^*}^1\right)
\frac{d\sigma}{d\cos\theta}\ d\cos\theta,
\end{eqnarray}
with $z^*>0$ such that 
\begin{equation}
\int_{-z^*}^{z^*}(1+\cos^2\theta)\ d\cos\theta
=\left(\int_{z^*}^1-\int_{-1}^{-z^*}\right)\ 2\cos\theta\ d\cos\theta.
\end{equation}
This condition implies same coefficients multiplying $F_1$ and $F_2$ after 
integration of Eq.~(\ref{cross}) over $\cos\theta$ in the indicated ranges, and 
determines the value 
of $z^*$ via a cubic equation with solution $z^*=2^{2/3}-1=0.5874$, 
corresponding to $\theta^*=54^\circ$. In the case of a reduced 
angular range, e.g., $\vert\cos\theta\vert<c$, one has ${z^*=(1+3c)^{1/3}-1}$. 
Also, one should notice that, in the approximation of neglecting $m_f$ as in 
(\ref{cross}), the value of $z^*$ is independent of $\sqrt s$. This would not 
be the case for ${\bar t}t$ pair production, where the corresponding 
$z^*$ would have a non-negligible dependence on $m_t/\sqrt s$, so that a 
separate treatment would be needed for this channel. In the case 
of Bhabha scattering, condition (8) could not be satisfied, 
and the corresponding $z^*$ could not be determined, 
because in this case the decomposition of the cross section into 
fully symmetric and antisymmetric parts as in (\ref{cross}) does not occur 
due to the presence of the additional $t$-channel $\gamma$- and $Z$-exchange 
amplitudes.
\par   
In terms of $F_1$ and $F_2$ of Eq.~(\ref{cross}), and of ${\sigma_{\rm
pt}\equiv\sigma(e^+e^-\to\gamma^\ast\to\mu^+\mu^-)}={
(4\pi\alpha_{e.m.}^2)/(3s)}$: 
\begin{equation}
\label{sigmapm}
\sigma_{\pm}=N_C\,\sigma^*_{\rm pt}\ (F_1\pm F_2),
\end{equation}
where, to a very good approximation: 
\begin{equation}
\sigma^*_{\rm pt}
=\frac{3}{4}\left(1-z^*{}^2\right)\sigma_{\rm pt}
=0.5\,\sigma_{\rm pt}.
\label{pt*}
\end{equation}
Introducing the helicity cross sections 
\begin{equation}
\label{hel}
\sigma_{\alpha\beta}=N_C\,\sigma_{\rm pt}\,\vert A_{\alpha\beta}\vert^2,
\end{equation}
and using Eqs.~(\ref{sigmapm}), (\ref{pt*}) and (\ref{f12}), the  
observables $\sigma_+$ and $\sigma_-$ can be expressed as:
\begin{equation}
\label{s+}
\sigma_+=\frac{1}{4}\,
\left[(1+P_e)(1- P_{\bar{e}})\,\sigma_{RR}
+(1-P_e)(1+P_{\bar{e}})\,\sigma_{LL}\right],
\end{equation}
\begin{equation}
\label{s-}
\sigma_-=\frac{1}{4}\,
\left[(1+P_e)(1- P_{\bar{e}})\,\sigma_{RL}
+(1-P_e)(1+ P_{\bar{e}})\,\sigma_{LR}\right].
\end{equation}
Eqs.~(\ref{s+}) and (\ref{s-}) show that $\sigma_+$ and $\sigma_-$ provide a 
convenient tool to separate cross sections with different 
combinations of helicities by different choices of the initial beams 
polarizations, and actually, in this regard,  they should be
interesting by themselves. Indeed, corresponding to the different initial
electron longitudinal right- and left-handed polarizations in 
Eq.~(\ref{f12}), one has: 
\begin{eqnarray} 
P_e=\pm 1,\hskip 4pt {P_{\bar{e}}}=0&:&\qquad\quad
\sigma_+^{R,L}\propto\sigma_{RR},\hskip 4pt \sigma_{LL}\nonumber\\ 
P_e=\pm1,\hskip 4pt { P_{\bar{e}}}=0&:&\qquad\quad \sigma_-^{R,L}\propto 
\sigma_{RL},\hskip 4pt \sigma_{LR}.\label{combi}\end{eqnarray}
For reference, we quote also the `conventional' observables for the analysis 
of process (\ref{proc}), namely, the total cross section
\begin{eqnarray}
\label{crosstot}
\sigma&=&\int\limits_{-1}^{1}
\frac{d\sigma}{d\cos\theta} d\cos\theta
=N_C\,\sigma_{\rm pt}F_1  \nonumber \\
&=&\frac{1}{4}\left[(1+P_e)(1- P_{\bar{e}})(\sigma_{RR}+\sigma_{RL})+
(1-P_e)(1+P_{\bar{e}})(\sigma_{LL}+\sigma_{LR})\right],
\end{eqnarray}
and the forward-backward asymmetry 
\begin{equation}
\label{AFB}
A_{\rm FB}=(\sigma^{\rm F}-\sigma^{\rm B})/
\sigma
=3F_2/4F_1,
\end{equation}
where 
$\sigma^{\rm F}
=\int_{0}^{1}(d\sigma/d\cos\theta)d\cos\theta$ and, similarly, 
${\sigma^{\rm B}
=\int_{-1}^{0}(d\sigma/d\cos\theta)d\cos\theta}$.
\par The independent observables $\sigma_+$ and $\sigma_-$
are simply related to the, also independent, $\sigma$ and $A_{\rm FB}$ by 
the relation 
\begin{equation}
\label{relation}
\sigma_{\pm}=0.5\,\sigma
\left(1\pm\frac{4}{3}A_{\rm FB}\right).
\end{equation}
Therefore, $\sigma_+$ and $\sigma_-$ can be measured either by direct 
integration of the differential cross section according to 
Eqs.~(\ref{sigma+}) and (\ref{sigma-}), or by the particular combination of
$\sigma$ and $A_{\rm FB}$ on the right-hand side of Eq.~(\ref{relation}), 
which carries the same kind of information.
\par 
From Eqs.~(\ref{s+}) and (\ref{s-}), only $\sigma_+$ and $\sigma_-$ allow to 
directly disentangle the helicity cross sections, by combining 
measurements at two different electron polarizations. From 
Eqs.~(\ref{crosstot}) and (\ref{AFB}), $\sigma$ and $A_{FB}$ give 
information only on linear combinations of helicity cross sections even 
for polarized electrons, and therefore do not allow such a direct separation 
by themselves.  The above mentioned distinctive features with regard to the 
determination of the helicity cross sections make 
$\sigma_+$ and $\sigma_-$ potentially more convenient, in order to study the 
deviations from the SM due to the four-fermion contact interactions. 
The role of these observables also for other types of new physics to be 
studied in 
$e^+e^-$ collisions such as, e.g., a new heavy neutral gauge boson 
$Z^\prime$ or the anomalous gauge boson couplings, has previously been 
emphasized in \cite{Osland}, \cite{Paver}.
\par  
The previous formulae continue to hold to a very good approximation 
with the inclusion of one-loop SM electroweak radiative corrections, in the 
form of improved Born amplitudes. Basically, the parameterisation that 
uses the   
best known SM parameters $G_{\rm F}$, $M_Z$ and $\alpha(M^2_Z)$ is obtained by 
the following replacements in the above 
equations \cite{Hollik}, \cite{Altarelli2}:
\begin{eqnarray}
\label{impborn}
\alpha_{e.m.}&\Rightarrow&\alpha_{e.m.}(M_Z^2) \nonumber
\\
g^f_L&\Rightarrow&
\frac{1}{\sqrt{\kappa}}
\left(I^f_{3L}-Q_f\sin^2\theta^{\rm eff}_W\right), \qquad
g^f_R\Rightarrow
-\frac{Q_f}{\sqrt{\kappa}}\sin^2\theta^{\rm eff}_W
\nonumber \\
\sin^2\theta_W
&\Rightarrow&
\sin^2\theta^{\rm eff}_W, \qquad
\sin^2(2\theta^{\rm eff}_W)
\equiv\kappa=\frac{4\pi\alpha(M_Z^2)}{\sqrt{2}G_{\rm F}\, M_Z^2\rho},
\end{eqnarray}
with 
\begin{equation}
\label{rho}
\rho\approx 1+\frac{3 G_{\rm F} m^2_{top}}{8\pi^2\sqrt{2}}.
\end{equation}
Moreover, for the $Z$-propagator: 
${\displaystyle{\chi_Z(s)\Rightarrow\frac{s}{s-M_Z^2+i(s/M_Z^2)M_Z\Gamma_Z}}}$.

\section{Model independent analysis and results}
According to Eq.~(\ref{combi}), by the measurements of $\sigma_+$ and 
$\sigma_-$ for the different initial electron beam polarizations one 
determines the cross sections $\sigma_{\alpha\beta}$ related to definite 
helicities. 
From Eq.~(\ref{amplit}), one can observe that the contact interaction 
contributes to these amplitudes the term 
${s\eta_{\alpha\beta}}/{2\alpha_{e.m.}\Lambda^2}$. To the same leading order in 
$s/\Lambda^2$, interference between the contact terms and the usual gauge 
interactions can affect the observables of process (\ref{proc}) and 
lead to deviations from the SM predictions at energies much below $\Lambda$ 
that in principle might be observed. The size of such interference term 
relative to the SM prediction is given by $s/\alpha_i\Lambda^2$, 
where $\alpha_i$ represents the strength of the relevant gauge coupling, and 
to this order one may neglect modifications of the gauge couplings due to form 
factors \cite{Godfrey}.
\par
Accordingly, in the considered situation $\sqrt s\ll \Lambda$, where only 
the interference term in the relevant observables is expected to be 
important, the deviation of each of the helicity cross sections from the SM 
prediction is given by the following expression:
\begin{equation}
\Delta\sigma_{\alpha\beta}\equiv 
\sigma_{\alpha\beta}-\sigma_{\alpha\beta}^{SM}= 
N_C\, \sigma_{\rm pt}\, 2 \, 
\left(Q_e\, Q_f+g_{\alpha}^e\, g_{\beta}^f\,\chi_Z\right)\cdot
\frac{s\eta_{\alpha\beta}}{2\alpha_{e.m.}\Lambda^2},
\label{deltasig}\end{equation}
and depends on a single `effective' non-standard 
parameter. Therefore, in an analysis of experimental data for
$\sigma_{\alpha\beta}$ based on a $\chi^2$ procedure, a one-parameter fit is 
involved and we may  hope to get slightly better sensitivity to 
contact interactions with respect to the other observables, i.e.
$\sigma$ and $A_{\rm FB}$, which depend on sums of different 
helicity cross sections and, consequently, involve more then one such free 
parameters at the same time. Moreover, in these cases, cancellations among the 
different independent parameters in interference terms cannot be excluded 
{\it a priori}. 
\par 
In the case where no deviations are observed, one can make an assessment 
of the sensitivity of process (\ref{proc}) to the contact interaction 
parameters, based on the expected experimental accuracy on the observables 
$\sigma_+$ and $\sigma_-$ introduced above. To this purpose, we adopt a 
$\chi^2$ procedure which starts from a 
$\chi^2$ function defined, for any observable $\cal O$, as follows: 
\begin{equation}
\label{Eq:chisq}
\chi^2
=\left(\frac{\Delta{\cal O}}{\delta{\cal O}}\right)^2.
\label{chi2}
\end{equation}
Here, $\delta{\cal O}$ is the expected uncertainty on the considered 
observable, and combines both statistical
and systematic uncertainties. As a criterion to constrain the values of 
the contact interaction parameters to the domain allowed by the 
non-observation of the corresponding deviations within $\delta\cal O$, we 
impose that $\chi^2<\chi^2_{\rm crit}$, where the actual value of 
$\chi^2_{\rm crit}$ specifies the desired `confidence' level. The numerical 
analysis has been performed by means of the program ZEFIT, adapted to the 
present discussion, which has to be used along with 
ZFITTER \cite{zfitter}, with input values $m_{top}=175$~GeV and $m_H=300$~GeV.
In order to reach the full sensitivity to contact interaction effects, a cut 
on the energy of photons emitted in the initial state, 
$\Delta=E_\gamma/E_{beam}$, is applied. For instance, at $\sqrt{s}=0.5$ TeV a 
radiative return to the $Z$ peak is avoided by choosing $\Delta=0.9$. 
\par
In practice, referring to Eq.~(\ref{combi}), polarization will not be exact, 
i.e., $\vert P_e\vert<1$. Therefore, the measured $\sigma_+$ of 
Eq.~(\ref{s+}) will involve a  linear combination of $\sigma_{LL}$ and 
$\sigma_{RR}$, which have to be disentangled from the data by solving the 
system of two equations corresponding to both signs of the electron 
longitudinal polarization, and the same is true for the determination of 
$\sigma_{RL}$ and $\sigma_{LR}$ from $\sigma_-$. For a quantitative 
discussion, we assume in the sequel $P_e=\pm P=\pm 0.8$ ($P_{\bar{e}}=0$) at 
the LC \cite{Accomando}. 
\par
For definiteness, we present in detail the case of $\sigma_+$.
The solutions of the system of two equations corresponding to $P_e=\pm P$ 
in Eq.~(\ref{s+}), can be written as:
\begin{equation}
\label{SRR}
\sigma_{RR}=\frac{1+P}{P}\sigma_{+}(P)-\frac{1-P}{P}\sigma_{+}(-P),
\end{equation}
\begin{equation}
\label{SLL}
\sigma_{LL}=\frac{1+P}{P}\sigma_{+}(-P)-\frac{1-P}{P}\sigma_{+}(P).
\end{equation}
From these equations, one can easily see that this procedure to 
extract $\sigma_{LL}$ and $\sigma_{RR}$, by the two independent measurements of 
$\sigma(\pm P)$, is efficient as long as the values of $\sigma_+(P)$ and 
$\sigma_+(-P)$, as well as their experimental uncertainties, are 
comparable. In general, the statistical uncertainty on a indirectly 
measured quantity such as, e.g., $\sigma_{RR}$ {\it via} $\sigma_{+}(P)$ 
and $\sigma_{+}(-P)$, is given by
\begin{equation}
\label{stat}
\delta\sigma_{RR}^{stat}=\sqrt{
\left(\frac{1+P}{P}\right)^2\left(\delta\sigma_{+}(P)\right)^2+
\left(\frac{1-P}{P}\right)^2\left(\delta\sigma_{+}(-P)\right)^2},
\end{equation}
where $\delta\sigma_{+}(\pm P)$ is the statistical uncertainty on 
$\sigma_{+}^{SM}(\pm P)$:
\begin{equation}
\label{dels+}
\delta\sigma_{+}(\pm P)=\sqrt{\frac{\sigma^{SM}(\pm P)}{\epsilon\, 
{\cal L}_{int}}}.
\end{equation}
Here, ${\cal L}_{int}$ is the integrated luminosity, $\epsilon$ is the 
efficiency for detecting the final state under consideration and 
$\sigma^{SM}(\pm P)$ is the polarized cross section defined by 
Eq.~(\ref{crosstot}). Eq.~(\ref{dels+}) has been obtained under the 
assumption that $\sigma_{+}^{SM}(\pm P)$ is measured directly as 
the difference of integrated cross sections defined in Eq.~(\ref{sigma+}).
Replacing Eq.~(\ref{dels+}) into Eq.~(\ref{stat}) one can easily 
find:\footnote{The same 
level of statistical uncertainty would obtain from the combination 
of $\sigma$ and $A_{FB}$ in Eq.~(\ref{relation}).} 
\begin{equation}
\label{statRR}
\delta\sigma_{RR}^{stat}=2\,\sqrt{\frac{\sigma_R^{SM}}
{\epsilon\, {\cal L}_{int}}
\left[1+\frac{1-P^2}{4P^2}\left(1+\frac{\sigma_L^{SM}}{\sigma_R^{SM}}
\right)\right]},
\end{equation}
where we introduced the following notations:
\begin{equation}
\label{sigRL}
\sigma_{R}^{SM}=\frac{1}{2}\left(\sigma_{RR}^{SM}+\sigma_{RL}^{SM}\right),
\quad
\sigma_{L}^{SM}=\frac{1}{2}\left(\sigma_{LL}^{SM}+\sigma_{LR}^{SM}\right).
\end{equation}
One should notice that the expression for $\delta\sigma_{RL}^{stat}$
is same as Eq.~(\ref{statRR}), whereas the expression for  
$\delta\sigma_{LL}^{stat}$ (=$\delta\sigma_{LR}^{stat}$) can be obtained 
from $\delta\sigma_{RR}^{stat}$ by changing $L\leftrightarrow R$.
The right-hand side of Eq.~(\ref{statRR}) has a non-trivial dependence on the 
value of the polarization $P$. In the numerical analysis presented below, 
we take three different values of the polarization, $P=$1, 0.8, 0.5, in order 
to test this dependence. 
\par
It turns out that, to a reasonable approximation, the sensitivity 
of $\sigma_\pm$ to the contact interaction parameters in the `linear' 
approximation where, as anticipated, only the interference term is taken into 
account in the observable quantities,\footnote{For consistency, including 
quadratic terms in $1/\Lambda^2$ would require consideration of the 
dimension 8 operators in the effective Lagrangian.} can be simply expressed 
by the bounds directly following from Eqs.~(\ref{deltasig}) and (\ref{chi2}): 
\begin{equation}
\label{bounds}
\Lambda^{-2}_{(\alpha\beta)}<\sqrt{\chi^2_{crit}}\, \frac{\alpha_{e.m.}}{s}\,
\frac{\delta\sigma_{\alpha\beta}}{N_C\sigma_{pt}\, 
\vert A_{\alpha\beta}^{SM}\vert}.
\end{equation}
\par
Numerically, we take into account for $\sigma_+$ and $\sigma_-$ the expected 
identification efficiencies \cite{Damerall} and the systematic uncertainties 
on the various fermionic final states, for which we assume: $\epsilon=100\%$ 
and $\delta^{sys}=0.5\%$ for leptons; $\epsilon=60\%$ and 
$\delta^{sys}=1\%$ for $b$ quarks; $\epsilon=35\%$ and $\delta^{sys}=1.5\%$ 
for $c$ quarks. Also, $\chi^2_{crit}=3.84$ as typical for 95\% C.L. with a 
one-parameter fit. 
The 95\% C.L. lower bounds on the mass scales $\Lambda$ relevant to the 
four pieces of the contact interaction (\ref{lagra}) are reported in 
Tables~1 and 2, corresponding to $\sqrt s=0.5$ TeV, 
${\cal L}_{int}=50\ fb^{-1}$, and $\sqrt s=1$ TeV, 
${\cal L}_{int}=100\ fb^{-1}$, respectively.
\begin{table}[t]
\centering
\caption{95\% C.L. model-independent compositeness search reach in TeV at 
$e^+e^-$ linear collider with $E_{c.m.}=0.5\, TeV$ and 
${\cal L}_{int}=50\,fb^{-1}$.}
\medskip
\begin{tabular}{|c|c|c|c|c|c|}
\hline
process & $P$ & $\Lambda_{RR}$ & $\Lambda_{LL}$ & $\Lambda_{RL}$ &
$\Lambda_{LR}$  
\\ \hline
$e^+e^-\to\mu^+\mu^-$ & 1.0 & 30.1 & 30.1 & 21.7 & 21.0
\\ \hline
$e^+e^-\to\mu^+\mu^-$ & 0.8 & 28.4 & 28.7 & 20.3 & 19.8
\\ \hline
$e^+e^-\to\mu^+\mu^-$ & 0.5 & 24.2 & 24.8 & 17.1 & 16.9
\\ \hline
\hline
$e^+e^-\to{\overline{b}}b$ & 1.0 & 34.8 & 32.8 & 26.2 & 17.4
\\ \hline
$e^+e^-\to{\overline{b}}b$ & 0.8 & 30.6 & 32.0 & 22.8 & 16.7
\\ \hline
$e^+e^-\to{\overline{b}}b$ & 0.5 & 23.8 & 29.5 & 17.6 & 14.7
\\ \hline
\hline
$e^+e^-\to{\overline{c}}c$ & 1.0 & 28.6 & 26.1 & 15.8 & 19.1
\\ \hline
$e^+e^-\to{\overline{c}}c$ & 0.8 & 27.1 & 25.6 & 14.4 & 18.2
\\ \hline
$e^+e^-\to{\overline{c}}c$ & 0.5 & 23.2 & 23.7 & 11.7 & 15.8
\\ \hline
\end{tabular}
\label{tab:tab1}
\end{table}
\begin{table}
\centering
\caption{95\% C.L. model-independent compositeness search reach in TeV at 
$e^+e^-$ linear collider with $E_{c.m.}=1\, TeV$ and 
${\cal L}_{int}=100\,fb^{-1}$.}
\medskip
\begin{tabular}{|c|c|c|c|c|c|}
\hline
process & $P$ & $\Lambda_{RR}$ & $\Lambda_{LL}$ & $\Lambda_{RL}$ &
$\Lambda_{LR}$  
\\ \hline
$e^+e^-\to\mu^+\mu^-$ & 1.0 & 51.4 & 51.6 & 36.9 & 35.6
\\ \hline
$e^+e^-\to\mu^+\mu^-$ & 0.8 & 48.4 & 48.9 & 34.5 & 33.6
\\ \hline
$e^+e^-\to\mu^+\mu^-$ & 0.5 & 40.9 & 42.0 & 29.0 & 28.6
\\ \hline
\hline
$e^+e^-\to{\overline{b}}b$ & 1.0 & 59.6 & 58.6 & 43.1 & 29.6
\\ \hline
$e^+e^-\to{\overline{b}}b$ & 0.8 & 52.2 & 56.9 & 37.6 & 28.5
\\ \hline
$e^+e^-\to{\overline{b}}b$ & 0.5 & 40.4 & 51.5 & 28.9 & 25.2
\\ \hline
\hline
$e^+e^-\to{\overline{c}}c$ & 1.0 & 51.1 & 47.9 & 27.3 & 32.5
\\ \hline
$e^+e^-\to{\overline{c}}c$ & 0.8 & 47.7 & 46.5 & 25.0 & 31.0
\\ \hline
$e^+e^-\to{\overline{c}}c$ & 0.5 & 40.0 & 42.1 & 20.3 & 27.0
\\ \hline
\end{tabular}
\label{tab:tab2}
\end{table}
Also, for polarized beams, we assume 1/2 of the total integrated luminosity
quoted above for each value of the electron polarization, $P_e=\pm P$.
Tables~1 and 2 show that the `new' integrated observables $\sigma_+$ and 
$\sigma_-$ are quite 
sensitive to contact interactions, with discovery limits ranging from 30 to 60 
times the CM energy at the maximal planned value of degree of the electron  
longitudinal polarization $P=0.8$. The best sensitivity occurs for the 
$\bar{b}b$ final state, while the worst one is for $\bar{c}c$.
Decreasing the electron polarization from $P=1$ to $P=0.5$ results in 
worsening the sensitivity by $20-30\%$, depending on the final channel, 
which is not dramatic. Regarding the role of the assumed uncertainties on the 
observables under consideration, in the cases of $\Lambda_{RL}$ and 
$\Lambda_{LR}$ the expected statistics are such that the uncertainty turns 
out to be dominated by the statistical one, and the results are almost 
insensitive to the value of the systematical uncertainty. Conversely, in the 
cases of $\Lambda_{LL}$ and $\Lambda_{RR}$ the results depend more 
sensitively on the chosen value of the systematic uncertainty. Moreover, one 
should remark that, as evident from Eqs.~(\ref{s+}) and (\ref{s-}), a further 
improvement on the sensitivity to the various $\Lambda$-scales in Tables 1 
and 2 would be obtained if both initial $e^-$ and $e^+$ longitudinal 
polarizations were available \cite{Accomando}.
\par
In conclusion, we have studied the sensitivity to four-fermion contact 
interaction effects at linear colliders of two `new' polarized 
observables, $\sigma_+$ and $\sigma_-$, leading to an analysis that enables 
to directly disentangle the four effective couplings relevant to the 
Lagrangian of Eq.~(\ref{lagra}). This feature can be realized by extracting 
individual helicity cross sections from the combination of observables 
measured at two different values of the electron polarization. Depending 
on the specific final state flavor and the helicity of fermions involved in 
process  (\ref{proc}), contact interactions can be probed up to vales of the 
corresponding mass scales $\Lambda$ of the order of  
$30-60$ times the CM energy.

\section*{Acknowledgements}
A.A. Pankov gratefully acknowledges the support and hospitality of the 
Theory Division of CERN, where part of this work was done. 
\vfill\eject

\end{document}